\documentclass[a4paper, 11pt]{article}
\usepackage[affil-it]{authblk} 
\usepackage{etoolbox}
\usepackage{lmodern}
\usepackage[a4paper]{geometry}
\usepackage[in]{fullpage}
\usepackage[utf8]{inputenc}
\usepackage{amsmath}
\usepackage{amssymb}
\usepackage{cite}

\usepackage{amsfonts}
\usepackage{graphicx}
\usepackage[dvipsnames]{xcolor}
\usepackage[colorlinks]{hyperref}
\hypersetup{linkcolor=blue,citecolor=blue,urlcolor=blue}

\title{On the TransPlanckian Censorship Conjecture in Holographic Cosmology}

\makeatletter
\patchcmd{\@maketitle}{\LARGE \@title}{\fontsize{16}{19.2}\selectfont\@title}{}{}
\makeatother

\author[1]{Heliudson Bernardo\footnote{Email: \href{mailto:heliudson@hep.physics.mcgill.ca}{heliudson@hep.physics.mcgill.ca}}}

\affil[1]{Department of Physics, McGill University,\protect\\ Montreal, QC, H3A 2T8, Canada}

\date{\vspace{-5ex}}

\begin{document}

\maketitle


\begin{abstract}
In this letter we discuss the implications of the recently proposed TransPlanckian Censorship Conjecture (TCC) for Holographic Cosmology. For models of holographic inflation that are dual to inverse renormalization group (RG) flows, assuming the TCC to be true, we found a bound on the running of the field theory coupling and the corresponding maximum energy scale above which the RG flow should be modified. We also discuss on the transPlanckian problem in non-geometrical holographic models.

\end{abstract}



\section{Introduction}
\label{sec:intro}

Over the last decades cosmological observations have been useful to constrain fundamental physics. Measurements of the cosmic microwave background radiation (CMBR) anisotropies \cite{Akrami:2018vks, Aghanim:2019ame} provide a unique way to test models of very early cosmology through the cosmological perturbation data that they would imprint in the sky. Most popular proposals to generate the spectrum of cosmological fluctuations are based on cosmological inflation \cite{Guth:1980zm, Linde:1981mu, Starobinsky:1980te}, though there are alternative models \cite{Brandenberger:2009jq}. According to inflation, the observed perturbation in the matter density that triggered the formation of large scale structure in the universe originated from vacuum fluctuations that exit the Hubble radius during inflation and then squeeze and classicalize before re-entering to produce the CMBR anisotropies. 

There are still various theoretical questions about inflation and its self-consistency (see \cite{Brandenberger:2012aj} and references therein). In particular, in \cite{Martin:2000xs} (see also \cite{Niemeyer:2000eh, Easther:2001fi}) it was shown that if inflation lasts slightly longer than the necessary to solve the Standard Cosmology shortcomings, then it would probe transPlanckian scales on which we are not suppose to believe the results anymore. More recently in \cite{Bedroya:2019snp} it was conjectured that this \emph{TransPlanckian problem} should not arise in scalar field based models of cosmology compatible with quantum gravity, what is called the \emph{TransPlanckian Censorship Conjecture} (TCC). The TCC states that quantum fluctuation modes of a field theory consistent with quantum gravity should never cross the Hubble radius and thus become classical in inflationary cosmology. Quantitatively, as the physical length scales are proportional to the scale factor $a(t)$, we have
\begin{equation}\label{tcc}
    \frac{a_f}{a_i} < \frac{M_{\text{Pl}}}{H_f},
\end{equation}
where $H_f$ is the Hubble expansion rate at $t_f$ when $a(t_f) = a_f$ and $a_i$ is the initial value of the scale factor at the beginning of inflation. 

\emph{If TCC is true} then a given tentative field theory that does not satisfy equation (\ref{tcc}) has no UV completion in String Theory. Indeed, the TCC was proposed as yet another criterion in the swampland program\footnote{See \cite{Saito:2019tkc} for a criticism on treating TCC as a Swampland criterion.}, a web of conjectures about String Theory that gained recent attention as a way to map the string landscape of effective field theories \cite{Ooguri:2006in,Obied:2018sgi, Garg:2018reu, Ooguri:2018wrx} (for reviews see \cite{Palti:2019pca, Brennan:2017rbf}). Relations between TCC and previous conjectures were explored in \cite{Bedroya:2019snp, Brahma:2019vpl}.  

While the Standard Cosmology evolution and the known alternatives for inflation are compatible with the TCC, its implications to inflationary cosmology were explored in \cite{Bedroya:2019tba} where it was found that generic models of inflation compatible with the TCC and with observations should have the slow-roll parameter $\epsilon$ satisfying $\epsilon < 10^{-31}$ and thus a very small tensor to scalar ratio $r< 10^{-30}$ (for ways to alleviate these bounds, see \cite{Mizuno:2019bxy,Brahma:2019unn, Berera:2019zdd, Cai:2019hge}) and for models of single-field inflation with canonically normalized scalar field it was shown that the only compatible models suffer from fine-tuning for the initial condition of the inflaton field. 

Notwithstanding such developments, it is now widely believed that gravity is holographic. Holography started to be seriously considered after the explicit String Theory construction of a correspondence between gravity in $D$-dimensional AdS spaces and conformal field theory (CFT) in flat $(D-1)$-dimensions \cite{Maldacena:1997re, Gubser:1998bc, Witten:1998qj}. Having several checks supporting gauge/gravity correspondences \cite{Aharony:1999ti} and having exploited how the holographic dictionary generically works \cite{Skenderis:2002wp, Papadimitriou:2004ap,Kanitscheider:2008kd} one could hope to apply holography for cosmology, that is, to describe cosmology in $D$-dimensions using a quantum field theory (QFT) in $(D-1)$-dimensions.

Attempts to construct QFT duals to Cosmology started to be explored not long after the first concrete proposal of the AdS/CFT correspondence was developed. The main goal of holographic cosmology is to describe the strongly coupled gravitational dynamics of very early universe by means of a dual field theory, allowing for instance the holographic calculation of cosmological perturbations. Since most popular models of early cosmology are based on inflation, the first theoretical explorations were focused on trying to construct a dS/CFT correspondence \cite{Strominger:2001pn, Witten:2001kn, Maldacena:2002vr}. This turns out to be very challenging, the first obstacle appearing from the fact that quantum correlators of scalar fields in AdS and dS are not mapped into each other under naive analytic continuation. In general, the difficulties to obtain an explicit example of dS/CFT comes from reconciling the dS symmetries with supersymmetry which translates into difficulties in constructing dS vacua in String Theory \cite{Danielsson:2018ztv}. 

In this letter we will be focusing on the more general approach of \cite{McFadden:2009fg, McFadden:2010na}, in which the holographic dictionary is built on early ideas of \cite{Witten:1998qj,Maldacena:2002vr} where it is assumed that the partition function of the QFT equals the wavefunction of the universe. The difference is that while in \cite{Maldacena:2002vr} there is an explicit analytic continuation on the dS radius, in \cite{McFadden:2010na} a correspondence between cosmological solutions and Euclidean domain walls is used such that it is Newton's constant and momenta that are continued instead. Holography for domain walls is well understood for asymptotic AdS and power-law solutions \cite{Skenderis:2002wp} and so it is possible to use previously known results for cosmological solutions under the domain wall/cosmology correspondence. There have been a lot of explorations of such classes of holographic models, in particular the scalar and tensor power spectra of cosmological perturbations were calculated holographically in \cite{McFadden:2010na} where, assuming a weakly coupled 3 dimensional superrenormalizable SU($N$) gauge theory dual to the domain wall, an alternative cosmological model was proposed in which the early universe is a non-geometrical strong gravitationally coupled phase (for an attempt to construct the dual field theory from String Theory see \cite{Bernardo:2018cow}). In \cite{Afshordi:2016dvb, Afshordi:2017ihr}, it was shown that this model fits the Planck data and that when compared with the $\Lambda$CDM the data has no preference between the models. More recently, in \cite{Nastase:2019rsn, Nastase:2019} it was shown how the horizon, flatness, relic and entropy problems are mapped to and can potentially be solved by the holographic non-geometrical model (for a discussion of the cosmological constant problem in such a framework, see \cite{Nastase:2018cbf}).

On the other hand, it seems that it is redundant to construct QFT duals for models in which the early phase is weak gravitationally coupled such as inflation, as it is possible to calculate the cosmological perturbations on the gravity side using the standard theory of cosmological perturbations \cite{Mukhanov:1990me} (see \cite{Brandenberger:2003vk} for a review). Moreover, the dual QFT should be non-perturbative in such models, and then it would be very difficult to get any useful information from the QFT side. Although these are valid arguments, it is important to notice that if we would like to link the non-geometrical phase with the usual known weak gravitationally coupled phase, it is better to have a holographic description of the latter, at least around the transition point. Nevertheless, we will focus on models in which though the QFT is strongly coupled, it has a UV critical point and the inflationary dynamics is dual to the inverse renormalization group (RG) flow induced after introducing a deformation in the CFT \cite{Strominger:2001gp, Balasubramanian:2001nb, Larsen:2002et, Halyo:2002zg, Larsen:2003pf, vanderSchaar:2003sz,Maldacena:2011nz, Hartle:2012tv, Bzowski:2012ih, Garriga:2013rpa, McFadden:2013ria, Garriga:2014fda, Isono:2016yyj, Achucarro:2018ngj}. In this case the RG flow is described by the breaking of conformal symmetry and it is possible to use \emph{conformal perturbation theory} to calculate the scalar and tensor power spectra of cosmological perturbations holographically.

Assuming the TransPlanckian Censorship Conjecture to be true, if a given cosmology is really holographically dual to a field theory with such a map constructed from String Theory, then the TCC should be derived from the properties of the dual theory. The purpose of this letter is to discuss the TCC constraints on the RG flow of the hypothetical dual field theory for cosmology. In the following section we translate the TCC bound (\ref{tcc}) to QFT language and then, focusing on models of holographic inflation that are dual to RG flows of scalar deformed CFT's, we show how the TCC gives a relation between properties of the scalar operator deformation and the scales in the RG flow. We find a maximum energy scale above which we cannot neglect extra deformations. In Section \ref{sec3}, we quickly discuss on the TCC for non-geometrical models and after that we conclude.

\section{TCC constraint on Holographic Inflation}\label{sec2}

The TCC can be directly translated in QFT language by recalling that cosmological time evolution is dual to inverse RG flow \cite{Strominger:2001gp, McFadden:2010na}. In fact, as already discussed in \cite{Nastase:2019rsn} the amount of RG flow between energy scales $q_i$ and $q_f$ is given exactly by the factor on the LHS of the inequality (\ref{tcc}). This can be seen from the relation between a fixed distance scale $L$ on the boundary and how deep a geodesic connecting points separated by such a distance probes the bulk. Due to the domain wall/cosmology correspondence, we will use results for domain walls to answer questions in cosmology. For quasi-de Sitter inflation, the relevant domain wall solution is asymptotically AdS, i.e.
\begin{equation}
    ds^2 = dz^2 + e^{2z}\delta_{ij}dx^i dx^j, \quad \text{as} \quad z\rightarrow \infty,
\end{equation}
which is just the Poincar\'e patch of Euclidean AdS. Following \cite{Nastase:2019rsn}, the geodesic distance calculation gives $L \propto e^{-z}$. Thus, by using the domain wall/cosmology correspondence we get the following relation between distance scales in the dual QFT
\begin{equation}
    \frac{a_f}{a_i} = \frac{L_i}{L_f},
\end{equation}
such that, as $L$ corresponds to a momentum scale $q \propto L^{-1}$, equation (\ref{tcc}) implies the following relation between the initial and final momentum change during the RG flow
\begin{equation}\label{tccholocosmo}
    \frac{q_f}{q_i} < \frac{M_{\text{Pl}}}{H(q_f)},
\end{equation}
with the Hubble radius dependence on the energy scale fixed by the QFT data and the holographic dictionary. 

As an illustrative example, for non-conformal D$p$-branes ($p \neq 3$), we would have $M^2_{\text{Pl}}/H^2 = \beta_p \sqrt{g_{\text{eff}}}$, where $\beta_p$ is a numerical factor that depends on $p$ and $g_{\text{eff}} = g_{\text{YM}}^2 N/ q^{3-p}$ is the dimensionful gauge coupling of the dual SU($N$) gauge theory with coupling $g_{\text{YM}}$ fixed by the normalization of the dilaton \cite{Kanitscheider:2008kd}. For the case relevant to holographic cosmology, $p=2$, equation (\ref{tccholocosmo}) can be written as
\begin{equation}
    \frac{q_f}{q_i} < \left(\frac{\beta_2^2g_{\text{YM}}^2 N}{q_f}\right)^{1/4},
\end{equation}
that implies an upper bound on the amount of inverse RG flow given by
\begin{equation}
    q_f < (\beta^2_2 g^2_{\text{YM}}N)^{1/5}q_i^{4/5}.
\end{equation}

For more concrete cases, in principle its is possible to constrain not only the amount but also the details of the RG flow such as the structure of beta functions. Focusing on models of holographic inflation where the QFT is a CFT deformed in the UV by a marginally relevant operator $\mathcal{O}_{\Delta}$ with dimension $\Delta = 3- \lambda$ and $\lambda \ll 1$, we have \cite{Strominger:2001gp, Larsen:2002et, Halyo:2002zg, vanderSchaar:2003sz, Bzowski:2012ih}
\begin{equation}
    S_{\text{QFT}} = S_{\text{CFT}} + \int d^3x \sqrt{g} \Lambda^{\lambda} \varphi\mathcal{O}_{\Delta}(x),
\end{equation}
with bare coupling $\varphi$ given at some energy scale $\Lambda$ close to the UV fixed point. The unperturbed two and three-point functions of $\mathcal{O}_{\Delta}$ can be written as
\begin{equation}
    \left\langle \mathcal{O}_{\Delta}(x_1) \mathcal{O}_{\Delta}(x_2)\right\rangle_0 = \frac{1}{|x_1 - x_2|^{2\Delta}}, \quad \left\langle \mathcal{O}_{\Delta}(x_1)\mathcal{O}_{\Delta}(x_2)\mathcal{O}_{\Delta}(x_3)\right\rangle_0 = \frac{C}{|x_1 - x_2|^{\Delta}|x_2 - x_3|^{\Delta}|x_3 - x_1|^{\Delta}},
\end{equation}
where the constant $C$ can be obtained from the operator product expansion (OPE)
\begin{equation}
    \mathcal{O}(x)\mathcal{O}(0) \sim \frac{1}{|x|^{2\Delta}} + \frac{C}{|x|^{\Delta}}\mathcal{O}(0) + \cdots
\end{equation}
Assuming that $C$ is a positive constant greater than unity, there is an IR fixed point that is close to the UV point, i.e., $\varphi$ is small during the flow and we can use conformal perturbation theory to calculate the correlators in the deformed theory. Explicitly, the beta function for $\varphi$ is (see for instance \cite{Bzowski:2012ih})
\begin{equation}\label{betafunction}
    \beta(\varphi) = \frac{d \varphi}{d \ln \Lambda} = -\lambda \varphi + 2\pi C \varphi^2 + O(\varphi^3),
\end{equation}
and so the IR fixed point is at
\begin{equation}
    \varphi = \frac{\lambda}{2\pi C} + O(\lambda^2).
\end{equation}
Solving equation (\ref{betafunction}), we get
\begin{equation}\label{phiintermsoflambda}
    \varphi = \frac{\varphi_1}{1 + \frac{\varphi_1}{g}\left(\frac{\Lambda}{\Lambda_0}\right)^{\lambda}},
\end{equation}
where $\varphi_1 = \lambda/2\pi C$ and $g$ is the renormalised coupling in the UV. The reference scale $\Lambda_0$ can be fixed by the choice of the normalization of $\mathcal{O}_{\Delta}$. 

The gravity dual to such a RG flow is a 4-dimensional domain wall solution with scalar potential (with $\varphi$ in Planckian units) \cite{Bzowski:2012ih}
\begin{equation}
    \kappa^2V_{\text{DW}}(\varphi) = H_{\text{UV}}^2\left(-3 + \frac{1}{2} \lambda(\lambda -3) \varphi^2 + 2\pi C(1-\lambda) \varphi^3\right) + O(\varphi^4),
\end{equation}
where $\kappa^{-2} = M^2_{\text{Pl}}$ and $H_{\text{UV}} \equiv -3\kappa^2V(\varphi =0)$. Under the domain wall/cosmology correspondence, this maps to a scalar field cosmology with potential $V(\varphi) = -V_{\text{DW}}(\varphi)$ which is a hill-top inflation potential. The equation of motion for the scalar field is simply \cite{Bzowski:2012ih}
\begin{equation}
    \frac{1}{H_{\text{UV}}}\frac{d\varphi}{dt} = -\lambda \varphi + 2\pi C \varphi^2 + O(\varphi^3),
\end{equation}
and, comparing with (\ref{betafunction}), we see that the time evolution may be identified with inverse RG flow, $H_{\text{UV}}t = \ln (\Lambda/\Lambda_0)$. The solution for the scale factor is \cite{Bzowski:2012ih}
\begin{equation}\label{scalefactor}
    a(t) = \left(\frac{g}{\varphi_1} + e^{\lambda H_{\text{UV}}t}\right)^{-\varphi_1^2/12}\exp{\left[H_{\text{UV}}\left(1+\frac{\lambda \varphi_1^2}{12}\right)t +\frac{1}{12}g\varphi_1 e^{\lambda H_{\text{UV}}t}\left(\frac{g}{\varphi_1} + e^{\lambda H_{\text{UV}}t}\right)^{-2}\right]},
\end{equation}
where $H_{\text{UV}} \equiv H(t\rightarrow \infty)$. Moreover, $H_{\text{IR}} \equiv H(t\rightarrow - \infty) = H_{\text{UV}} + \lambda\varphi_1^2/12$. These asymptotic values of $H(t)$ are defined by the data of the CFT at the critical points. Notice that $H_{\text{IR}}$ being greater than $H_{\text{UV}}$ was to be expected as a consequence of the c-theorem for RG flows \cite{Strominger:2001gp, Balasubramanian:2001nb}. Since the difference between the $H_{\text{IR}}$ and $H_{\text{UV}}$ is of order $O(\lambda^3)$, we can safely assume $H \simeq H_{\text{IR}}$ throughout the $\varphi$ evolution.

Having discussed the RG flow and its dual cosmology, we can now proceed to find how TCC constrains the RG flow. Focusing on the inflationary phase in the IR around $\varphi_1$, the $\epsilon$ slow-roll parameter is given by
\begin{equation}
    \epsilon \simeq \frac{1}{2} \lambda^2 \varphi^2\left(1 - \frac{\varphi}{\varphi_1}\right)^2,
\end{equation}
and, since $0 < \varphi < \varphi_1 \ll 1$, we can approximate $\epsilon \simeq \lambda^2 \varphi^2/2$. Thus the number of e-foldings is
\begin{equation}
    N \simeq \int \frac{1}{\sqrt{2\varepsilon}} d\varphi \simeq - \frac{1}{\lambda}\ln \varphi,
\end{equation}
and so the TCC bound (\ref{tcc}) implies
\begin{equation}
    \left(\frac{\varphi_i}{\varphi_f}\right)^{1/\lambda} < \frac{M_{\text{Pl}}}{H_{\text{IR}}},
\end{equation}
which gives a bound on the coupling of $\mathcal{O}_{\Delta}$,
\begin{equation}\label{boundonphi}
    \varphi_f > \frac{\lambda}{2\pi C}\left(\frac{H_{\text{IR}}}{M_{\text{Pl}}}\right)^{\lambda}. 
\end{equation}
Thus we see that the $\varphi$ is bounded from below by a combination of the dimension of the operator, its OPE coefficient $C$ and the IR scale of the CFT. Notice that upon using equation (\ref{phiintermsoflambda}) to write $\varphi_f$ in terms of the final energy scale $\Lambda_f$, the relation (\ref{boundonphi}) is compatible with (\ref{tccholocosmo}). Therefore, any realistic construction of holographic inflation that satisfies TCC should include extra field operators that should dominate the RG flow at $\Lambda_f$ whose maximum value is 
\begin{equation}
    \Lambda_f^{\text{max}} = \frac{M_{\text{Pl}}}{H_{\text{IR}}}\left(g \Lambda_0^{\lambda}\frac{2\pi C}{\lambda}\right)^{1/\lambda}.
\end{equation}
From equation (\ref{phiintermsoflambda}) the combination $g\Lambda_0^{\lambda}$ is the dimensionful coupling constant in the UV CFT.

Notice that being a hill-top inflation potential, this is an example of the ones considered in \cite{Bedroya:2019tba} and so their constraints are applicable in the present case. Since we have fixed $\varphi_i \simeq \varphi_1$, imposing the observational constraints on the power spectrum of fluctuations would relate $\varphi_f$ to the value of the potential or, equivalently, to $H_{\text{IR}}$. Choosing $\Lambda_0$ to match the pivot scale for the scalar power spectrum as in \cite{Bzowski:2012ih}, 
\begin{equation}
    \Lambda_0 = q_0 \left(\frac{\varphi_1}{g}\right)^{1/\lambda},
\end{equation}
the amplitude of the scalar power spectrum is 
\begin{equation}
    \Delta_S(q_0) = \frac{H^2_{\text{IR}}}{8\pi^2M^2_{\text{Pl}}\epsilon} \simeq \frac{H^2_{\text{IR}}}{4\pi^2M^2_{\text{Pl}}\lambda^2\varphi_1^2},
\end{equation}
and so using $\Delta_S \simeq 2\times 10^{-9}$ we get
\begin{equation}\label{Vfrompowerspectrum}
    \frac{V_0}{M^4_{\text{Pl}}} \simeq 2\times 10^{-9} 12 \pi^2 \lambda^2 \varphi_1^2. 
\end{equation}
As the second slow-roll parameter is given by $\eta \simeq -2\lambda$ we have $\epsilon \sim \lambda^4C^{-2} \ll \eta$ and the spectral index $n_s$ satisfy $1-n_s \simeq 2\lambda$. Thus $\lambda$ is fixed by observations to be $\lambda \simeq 0.02$ and equation (\ref{Vfrompowerspectrum}) relates $C$ with the scale of inflation,
\begin{equation}
    C \simeq 3.1\times 10^{-8} \sqrt{\frac{M^4_{\text{Pl}}}{V_0}}.
\end{equation}
Plugging this value into equation (\ref{boundonphi}) we find
\begin{equation}
    \varphi_f > 1.78 \times 10^5 \left(\frac{H_{\text{IR}}}{M_\text{Pl}}\right)^{1.02}.
\end{equation}
Taking $V_0^{1/4} = 10^{-10}M_{\text{Pl}}$ as in \cite{Bedroya:2019tba}, we get $C \simeq 3.1\times 10^{12}$, $\varphi_1 \simeq 1.03\times 10^{-15}$ and so for any model with such scale TCC is satisfied if $\varphi_f> 4.04 \times 10^{-16}$. 

\section{Discussion}\label{sec3}

From the solution (\ref{scalefactor}), it is possible to show that the time range for the coupling to undergo the entire inverse RG flow (from $\Lambda = 0$ to $\Lambda \to \infty$) is infinite. So the entire evolution through the ``thick" domain wall is inconsistent with TCC. This fact is not a problem for holographic inflation, as inflation happens close to the IR region. Furthermore we could have started with a 3 dimensional deformed CFT with a marginally irrelevant operator with dimension $\Delta = 3+\lambda$ in the IR and then we would get the same results. Thus, the fact that the entire RG flow is too ``long" does not invalidate the discussion in the previous section. 

A few words about the non-geometric holographic models are in place. Being alternatives to inflation, models of holographic cosmology in which gravity is strongly coupled such as \cite{McFadden:2013ria} are in principle not affected by TCC. In this case, the super-Hubble modes of cosmological perturbation emerge from the non-geometrical phase through the transition between such period to usual Standard Cosmology and their spectra can be calculated from correlators in the dual field theory, that is local and well defined. As stressed in \cite{Nastase:2019rsn}, the fact that the holographic dictionary is non-local explains how we could have correlation between super-Hubble scales even if there were no metric to define ``causality" during the non-geometric phase. 

It is worth mentioning that in such non-geometrical models gravity is becoming increasingly non-perturbative, due to the generalized conformal structure of the dual QFT \cite{McFadden:2013ria} and so in realistic models one should reverse this trend such that gravity becomes weak during the transition to Standard Cosmology. In this scenario, TCC could constraint such \emph{``holographic reheating"}. Regardless, one can continue the relation (\ref{tccholocosmo}) to the non-geometric phase in this class of models, in a similar fashion as the authors of \cite{Nastase:2019rsn} imposed the condition on the amount of RG flow in order to solve the flatness and horizon problems. In this case we would have a condition on the dual theory and \emph{once} we have an explicit construction of a dual pair from String Theory, it would be interesting to investigate what this condition implies to or if it even can be derived from the stringy construction.

\section{Conclusion}\label{sec6}
We have shown how the TransPlanckian Censorship Conjecture translates into a bound on the renormalization group flow of field theories of holographic cosmology. For models of holographic inflation in which the cosmological evolution is mapped to domain wall solutions that are dual to RG flows induced by marginal scalar deformations of conformal field theories, we found a bound on the coupling of the scalar operator that relates the energy scale of the theory, the dimension of the operator and one of its OPE coefficients. This bound corresponds to a maximum energy scale above we should include the contribution of extra deformations to the QFT. We also discussed on how the TransPlanckian problem of inflation could be solved in non-geometrical holographic cosmological models.

\section*{Acknowledgments}
The author is grateful to Horatiu Nastase and Kostas Skenderis for discussions about holographic cosmology and Robert Brandenberger, Suddhasattwa Brahma and Jéssica Martins for comments on the first version of the manuscript. The author's research is supported by funds from NSERC.




\bibliographystyle{JHEP} 
\bibliography{References}





\end{document}